\documentstyle[epsfig,aps]{revtex}
\newcommand{\ndt}{\noindent}

\newcommand{\ov}{\overline}

\begin{document}

%\begin{flushright}
%Bologna, 03/01/1999
%\end{flushright}

\begin{center}
{\bf\large Recent OPAL measurements in non-perturbative QCD }
\end{center}

\vspace{3mm}
\begin{center}
G. Giacomelli \par
Dipartimento di Fisica dell'Universita' di Bologna and INFN, Sezione di
Bologna, \\
Viale B. Pichat 6/2, 40127 Bologna, Italy \par
\vskip 0.3 cm
for the OPAL Collaboration \par
\end {center}
\vskip 0.5 cm

\ndt {\small {\bf Abstract.} Using 
multihadronic $Z^{0}$ decays recorded in
1991 - 1995 by the OPAL detector in $e^{+}e^{-}$ collisions at LEP1, 
experimental analyses were made of the following subjects:
(a) Bose-Einstein correlations; (b) Intermittency and correlations; 
(c) $ \rho$ and
$ \omega$ spin alignments; (d) A search for the tensor glueball candidate 
$f_{j}$(2200).}\par

\vspace{0.5cm}

\ndt {\bf 1 ~Introduction} \par~\par
  The large sample of multihadronic (MH) $ Z^{0}$ decays recorded by the OPAL 
detector at 
LEP1 offers the possibility to study several aspects of non 
perturbative QCD.

  Bose-Einstein correlations (BEC) give an estimate of the size and shape of the
identical bosons emitting region; more recently BEC have been used to yield 
information on the hadronization process.

  Intermittency and multiparticle correlations are studied in terms of the 
scaled normalized factorial moments and of cumulants vs rapidity $y$, 
azimuthal angle $ \phi$, and transverse momentum $P_{t}$. Intermittent
behaviour is seen in $y$ and $ \phi$ as an increase of the moments 
with decreasing bin size.
Genuine multiparticle correlations exist. They
are contained in the used Monte Carlos.

  Little is known on the role of spin in the hadronization process.
At LEP, this can be investigated by studying the properties of vector mesons
produced in hadronic $ Z^{0}$ decays. New results are presented on the helicity
density matrix elements for the $ \rho (770)^{\pm}$ 
and $ \omega (782)$ mesons.

  QCD predicts the existence of glueballs; lattice gauge theories predict the 
mass of the ground state glueball (two leading gluons with antiparallel spins)
to be $ \simeq$ 1.5 GeV. Some glueball candidates have been reported; 
the interpretation of data in this mass region is difficult because of
mixing with conventional mesons. This difficulty is less pronounced for the 
tensor glueball, expected with $ m \simeq 2.2$ GeV. A search for this state 
is here reported.

  The OPAL data at LEP1 concern 4.1 million MH 
events.
The analyses use charged particles  detected by the tracking chambers in the
magnetic field. Standard quality cuts are applied to each track and to each 
multiparticle event. Photons are detected in the  electromagnetic
calorimeter, or as $ \gamma \rightarrow e^{+} e^{-}$.\par
\vspace {0.5cm}     
  
\ndt {\bf 2 ~Bose-Einstein Correlations}\par~\par
Bose-Einstein correlations (BEC) between identical bosons lead to an 
enhancement of the number of identical bosons when they are close in phase 
space. LEP1 data allowed to study: (i) correlations 
for $\pi^{\pm} \pi^{\pm}$, $K^{0}K^{0}$, $K^{\pm}K^{\pm}$,
(ii) the multiplicity dependence of the $2\pi$ emitting
region, (iii) multi-pion BECs, (iv) two-dimensional $\pi\pi$ BEC [1]. 
LEP2 data allowed to study (v) BEC in 
$e^{+} e^{-} \rightarrow W^{+} W^{-}$ for identical
charged pion pairs. For two pions we use the notation in terms of
$Q^{2}= -(P_{1}-P_{2})^{2} = m^{2}_{\pi \pi}-4m^{2}_{\pi}$.
  The distribution in Q of the like-charge pairs has to be normalized to a 
Q-distribution which does not contain BEC, for instance the distribution of
unlike-charge pairs or to a Monte Carlo prediction [2]:
\begin{equation}
C^{\prime}(Q)=(N^{++}+N^{--})/N^{+-}~~,~~ 
C(Q)=(N^{++}+N^{--})/(N^{++}+N^{--})_{MC}
\label{eq:1}
\end{equation}
Various parametrizations are used in the literature to interpret the data. We 
use the Goldhaber parametrization:
\begin{equation}
 C(Q)= N (1+\lambda ~e^{-Q^{2} R^{2}})(1+\epsilon Q+\delta Q^{2})
\label{eq:2}
\end{equation}
\ndt N is a normalization factor, $ \lambda$ 
is the chaoticity parameter $ (0 \leq \lambda \leq 1)$  
and $R$ estimates the radius of the emitting
region; $(1+\epsilon Q+\delta Q^{2})$ 
accounts for the long range correlations arising form 
charge and energy conservation and phase space constraints.
The Q-dependence, corrected for Coulomb effects, of $C(Q)$ or $C^{\prime}(Q)$
 for $ \pi^{\pm} \pi^{\pm}$
shows an enhancement at small $Q$ value. Fits of  
$C(Q)$ to Eq. 2, eliminating the regions where there
 could be hadron 
resonance effects, yields $R\simeq 0.793\pm 0.007_{st} \pm 0.015_{sys}$ and
$\lambda \sim 0.58$ [2]; $R$ and $\lambda$ seem to be
energy independent [1].\par For $2 \pi$, OPAL analyzed the dependence of $R$ on 
the average charged multiplicity of MH events, and it found an
increase of $R$ with $n_{ch}$ (from 0.86 to 1.0 $fm$)
 and a decrease in the parameter $\lambda$ [3].
Similar values of $R$ and $\lambda$ were found for $\pi^{0} \pi^{0}$ [3],
$K_{s}^{0} K_{s}^{0}$ [4] and $K^{\pm} K^{\pm}$ [5] BECs.
For $3\pi$ BECs, OPAL [6] considered the ratio
\begin{equation}
R_{3}(Q_{3})=N^{+++}(Q_{3}) / N_{MC}^{+++}(Q_{3})
\label{eq:3}
\end{equation}
\ndt corrected for Coulomb effects; 
the Monte Carlo is JETSET7.4. The effects of 
$ 2 \pi$ 
correlations must be removed to obtain the genuine $ 3 \pi$ 
correlation function:
\begin{equation}
C_{3} (Q_{3})= R_{3}(Q_{3}) - R_{1,2}(Q_{3})=\frac{N^{+++}}{N_{MC}^{+++}} - 
\frac{\delta ^{++-}} {N_{MC}^{+++}}~\langle \frac{n^{+++}}{n^{++-}} \rangle
\label{eq:4}
\end{equation}
\ndt $ R_{1,2} = \delta _{2} ^{+++}/N_{MC}^{+++}$ ,~ $\delta _{2} ^{+++}$
 is the three-pion enhancement due to the two-pion BEC;
 $Q_{3}^{2}=-(q_{1}-q_{2})^{2} - (q_{1}-q_{3})^{2} - (q_{2}-q_{3})^{2}$
\par
The experimental distribution (4) was fitted to an expression similar to (2). 
%\begin{equation}
%C_{3}(Q_{3}) = N(1+2\lambda_{3} e^{-Q_{3}^{2} R_{3}^{2}})
 %(1+\epsilon Q_{3}+\delta Q_{3}^{2})
%\label{eq:5}
%\end{equation}
One obtains:
$ R_{3}=0.580 \pm 0.004_{st} \pm 0.029_{sys}~fm ~;
~\lambda_{3}=0.504 \pm 0.010_{st} \pm 0.041_{sys}$
Notice that two-pion BEC yield
$ R_{2}=0.793 \pm 0.015 ~fm $; assuming that
 $ R_{3} \simeq R_{2} / \sqrt{2}=0.793/ \sqrt{2} $ one has 
$ R_{3}\simeq 0.561 \pm 0.011 ~fm $ which
agreees with the measured $ R_{3}=0.558~ fm $.\par
BECs have been measured for $ 2 \pi $ in the reaction 
$ e^{+} e^{-} \rightarrow W^{+} W^{-}$ at 
$ \sqrt{s}=172$
and 183 GeV [7].
Three data samples are available: (i) the four-jet hadronic
sample, $ W^{+} W^{-} \rightarrow q \ov q \prime~ q \ov q \prime $: 
here final state interactions (FSI) can occur within a single $W$ 
decay or between
 the two different $W$ bosons.
(ii) The two-jet (semileptonic) sample,
$ W^{+} W^{-} \rightarrow q \ov q \prime~l \ov \nu_{l} $; FSI can only
occur within a single $W$. 
(iii) The leptonic sample, 
$ W^{+} W^{-} \rightarrow l^{+} \nu_{l}~ l^{-} \ov \nu_{l} $. 
From the point of view of the $M_{W}$ measurements, FSI are of consequence  if
they occur between the two $W$'s in the same hadronic event. At LEP2 
energies the two
$W$-bosons decay within a distance of about $0.1~ fm$, which is smaller than the
hadronization scale of $ \simeq 1~ fm$. Thus the fragmentation processes 
for hadrons 
coming from different $W$-bosons could be interconnected. 
OPAL demonstrated the existence of BEC in $W$ decays, Fig. 1. An 
attempt was made to determine BEC for two pions originating from the same
$W$-boson and from different $W$-bosons, as well as for pions from 
$ (Z^{0}/ \gamma) \rightarrow q \ov q $
events [7]. 
Fitting these samples together, assuming a common source radius $R$, we find
$ \lambda ^{same}=0.63 \pm 0.19 \pm 0.14~~,
~~\lambda ^{diff}= 0.22 \pm 0.53 \pm 0.14~~,
~~\lambda ^{Z^{\ast}}= 0.47 \pm 0.11 \pm 0.08~~,
~~R= 0.92 \pm 0.09 \pm 0.09 ~fm $
\ndt (the first error is statistical and the second one is systematic). 
At present
it is not established whether
BECs between pions from different $W$ bosons exist or not.\par
Two-dimensional, two-pion BECs are being studied;  preliminary
 results indicate a non 
spherical di-pion emitter shape [8].
\par
\vspace{0.5cm}
\ndt {\bf 3 ~Intermittency and multiparticle correlations}\par~\par
 Local multiparticle fluctuations have been studied in final hadron states. 
With the word Intermittency one refers to the rise of the
factorial moments with decreasing bin size. Analyses in terms of the
factorial cumulant moments reveal ``genuine multiparticle" correlations. 
The modified factorial moments (FM)
\begin{equation}
F_{q}^{c}=N ^{q} \overline{\langle n_{m}^{[q]} \rangle} / \overline{N_{m}^{[q]}}
\label{eq:6}
\end{equation}
\ndt are used by OPAL to avoid biases in the normalization [9]. 
The q-th order factorial moment is
$ n_{m}^{q}=n_{m} (n_{m}-1)...(n_{m}-q+1)$; $ n_{m} $ is the number of 
particles in the m-th bin of the phase space divided into $M$ equal bins,
$ N_{m} $ is 
the number of particles summed over all the N events, 
$ N_{m}= \sum_{i=1}^{n} (n_{m})_{i}$
 The bar stands
for ``horizontal" averaging over the bins in each event, 
$ (1/M) \sum_{m=1}^{M} $; 
the angle
brackets denote ``vertical" averaging over the events. The phase 
space is considered in terms of the rapidity $y$, azimuthal angle 
$ \phi$ and transverse
momentum $P_t$. Non-statistical fluctuations lead to increasing FMs with 
increasing $M$, $ F_{q} (M)\approx M^{\varphi_q} $, 
$ 0 ~< \varphi_{q}~ < q-1 $, $ M \rightarrow \infty $
(power dependence, scaling
law). The FMs of order $ q=2 \div 5 $ have been plotted 
vs the number of bins $M$ 
for the one-dimensional 
rapidity, azimuthal angle, and transverse momentum. The $y$ 
and $ \phi $ distributions increase
with $M$ up to $ M \sim 20$ and then flatten; 
the $ P_{t} $ distribution is essentially 
independent of $M$. Thus there is an intermittency in $y$ and 
$ \phi $ and not in 
$ P_{t} $. 
This behaviour is reasonably well predicted by the used Monte Carlos. The same 
behaviour, with stronger increase with $M$, is exhibited in two phase space 
dimensions, in particular by 
$ y \times \phi $, and in three dimensions by 
$ y \times \phi \times P_{t}$. 
  The modified factorial cumulant (FC) moments are
\begin{equation}
K_{q}^{c} = N \ov K_{q}^{(m)}~/~ \overline {N_{m}^{[q]}}
\label {eq:7}
\end {equation}
The multipliers $ K_{q}^{(m)} $
 are the unnormalized factorial cumulants (Mueller moments),
and represent genuine q-particle correlations. This leads to
\begin{equation}
F_{2} = K_{2} +1~~,~ F_{3} = K_{3} +3 K_{2}+1~~,
~ F_{4}= K_{4}+4K_{3}+3 \overline{(K_{2}^{(m)})^{2}} +6K_{2} +1~, ...
\label {eq:8}
\end {equation}
Two-particle contributions $ F_{q}^{(2)} $  can be expressed as
$F_{3}^{(2)}= 3 K_{2}+)1,~F_{4}^{(2)}= \ov{3(K_{2}^{(m)})^{2}} 
+6K_{2} +1,
~ F_{5}^{(2)}= \ov{15 (K_{2}^{(m)})^{2}}+ 10 K_{2} +1 $, ...
These equations can be used to search for multiparticle correlation 
contributions to local fluctuations. 
Fig. 2 shows the FCs in one, two and three dimensions for $ q=4 $ and 5.
the FCs are positive, which indicates that multiparticle 
correlations exist. The FCs exhibit stronger power-law behaviour than for FMs.
Addition of phase space dimensions, from 
$y$ to $ y \times \phi $ and to $ y \times \phi \times P_{t} $,
enhances the manifestation of multiparticle genuine correlatons [9]. 
The results agree with the QCD jet formation dynamics, but additional 
contributions from other mechanisms cannot be excluded.\par
\vspace{0.3cm}
\ndt {\bf 4 ~Spin 
alignment of {\boldmath $ \rho(770)^{\pm}$} and {\boldmath $ \omega (782)$}
 mesons}\par
\vspace{0.3cm}
The helicity density matrix elements for the $ \rho(770)^{\pm}$ and 
$ \omega (782)$ mesons
produced in $ Z^{0}$ decays have been measured 
by OPAL [10].
Since 
$ \rho^{\pm} \rightarrow \pi^{\pm} \pi ^{0} $,
 we can use as a spin analyzer the angle in the $\pi^{\pm}\pi ^{0} $,
 rest frame 
between one of the pion momenta and the $ \rho ^{\pm}$ boost direction. 
The distribution
of this angle, $\theta _{H}$, is:
\begin {equation}
W (\cos \theta _{H})= \frac {3}{4} [(1- \rho_{00})+(3 \rho_{00}-1)\cos ^{2} 
\theta _{H}] \label{eq:9}
\end {equation}
$ \rho_{00}$ is the helicity density matrix element corresponding to the 
probability that the spin of the $ \rho^{\pm}$ 
meson be perpendicular to its momentum
direction. Because of unitarity and parity conservation, the probabilities that
the spin be parallel or antiparallel, $ \rho_{11}$ 
and $ \rho _{-1-1}$, are equal  
$ \rho_{11}= \rho _{-1-1}= (1- \rho_{00})/2$.\par 
The $ \pi ^{\pm}$ and $ \pi^{0}$
candidates are combined in pairs and three quantities are evaluated: the
scaled energy of the pair $ x_{E}= E_{meson}/E_{beam}$, 
the invariant mass $ m_{\pi ^{\pm} \pi ^{0}}$, and $ \cos \theta _{H}$
defined between the $ \pi^{0}$ and the boost of the 
$ \pi^{\pm} \pi ^{0} $ system. The $ \rho^{\pm}$ signal in
the $ m_{\pi ^{\pm} \pi ^{0}}$ 
distribution is parametrized with a Breit-Wigner convoluted with the
experimental mass resolution in $ x_{E}$.
The efficiency-corrected 
$ \rho ^{\pm}$ yields are evaluated
for ten $ \cos \theta _{H}$ bins and fitted to the formula: 
$I(\cos \theta _{H})=A(1+B \cos^{2} \theta _{H})$.
\ndt One has $ \rho_{00}= (1+B)/(3+B) $. 
The parameters $A$, $B$ are obtained from a linear least
square fit.
  The decay $ \omega \rightarrow \pi ^{0} \pi ^{+} \pi^{-}$ 
has $BR=88.8 \% $. In the rest frame of the 
$ \pi ^{0} \pi ^{+} \pi^{-}$ system the
momenta of the three pions lie in a plane; the appropriate spin analyzer 
is the angle $ \theta _{H}$ between the normal to this plane and the 
boost direction. 
  The measured values of $ \rho_{00}$ as a function of $ x_{E}$ 
are shown in Figs. 3a and 3b
for the $ \rho^{\pm} $  and $ \omega$ mesons, respectively. 
The measurements are compatible 
with 1/3, corresponding to a statistical mix of helicity $ -1,0$ and $+1$ 
states.
The DELPHI results for $ \rho^{0}$ agree with our $ \rho^{\pm}$ values [11].
For $0.3 < x_{E}< 0.6$ one has 
$ \rho_{00} = 0.373 \pm 0.052$ 
and $ 0.142 \pm 0.114$ for $ \rho^{\pm}$ and $ \omega$, 
respectively. These results are lower than those for 
$ K^{\star}(890)$
and $\phi$ mesons [12]. Fig. 3c shows a comparison with previous data.\par
According to the Monte Carlo, up to $10 \% $ of $ \rho^{\pm}$ 
originate from decays of $ J^{P}=0^{-} $
mesons into a $ \rho^{\pm}~ (J^{P}=1^{-})$ 
and another $ J^{P}=0^{-} $ meson (mainly $ D^{0} \rightarrow \rho^{+} K^{-},~
J^{P} =0^{-} \rightarrow 0^{-}+1^{-})$. The effects of the
alignment of the $ \rho^{\pm}$ 
from such decays are shown in Fig. 3a as 
dotted lines.\par 
In conclusion there appears to be a difference in the spin alignment
properties of vector mesons, see Fig. 4, possibly depending on their 
strangeness content.\par
\vspace{0.3cm}
\ndt {\bf 5 ~A search for the tensor glueball 
{\boldmath $ f_{j}(2220)$}}\par
\vspace{0.3cm} 
In the $ K_{s}^{0}~K_{s}^{0}$ and $K^{+}~K^{-}$ 
final states a narrow resonance with $ \Gamma \simeq 20$ MeV 
(denoted $ f_{j}(2220)$ was reported
by the MARK III Coll., in $ J/ \psi$ decays [12], 
by the BES Collaboration
in radiative $ J/ \psi$ decays [13] 
and by the L3 Coll. [14] in hadronic $ Z^{0}$
decays in an enriched three-jet sample (and mainly coming from the lowest
energy jet, enriched in gluons). But the DM2 Coll., in $J/ \psi$ decays
[15] and the CLEO Coll., in $ \gamma \gamma$ 
collisions [16], did not find indications
for this state (the DM2 Coll. had indications for a broader structure). 
  The OPAL detector is well suited for a study of the 
$f_{j}(2220) \rightarrow K_{s}^{0}~K_{s}^{0}$ because of 
its large radius tracking chamber which allows efficient detection 
of $K_{s}^{0}$
over a large momentum range [17]. The $K_{s}^{0}~K_{s}^{0}$ 
mass spectrum was investigated in the full
event sample and, following the L3 analysis, in the three-jet sample and in 
individual jets of the three-jet sample.
Neither in the global event sample nor in gluon enriched samples is there any
evidence for a narrow resonance (see Fig. 5). 
The upper limit for the product of the
inclusive production rate per hadronic $ Z^{0}$ decay 
$ n_{f_{J}(2220)}$ and the branching ratio
$ BR_{f_{J} \rightarrow K \ov K}$ 
is $ [n_{f_{J}} (2220)~ \cdot ~  BR_{f_{J} \rightarrow K \ov K}]<0.0028$, 
at $95 \% $ C.L. for the $f_{j}(2220)$ with
production and decay properties as a $ 2^{++}$ meson.\par
\vspace{0.3cm}
\ndt {\bf 6 ~Conclusions}\par
\vspace{0.3cm}
Many measurements related to multihadronic non perturbative QCD have been 
performed by OPAL. In particular:
(i) The studies of BEC is yielding a number of interesting informations;
(ii) Intermittency and multiparticle correlations are present in multihadron
events;
(iii) More precise experiments are needed to understand the role of
spin in the hadronization process;
(iv) OPAL found no confirmation for the proposed $ f_{j}(2220)$ tensor glueball 
candidate.
\vspace{0.5cm} \par
\ndt {\bf References}\par~\par
\ndt [1] E. A. De Wolf, Bose-Einstein Correlations, Int. Conf. High
En. Phys., Glasgow (1994), Inst. Phys. Publ. \par 
(1995), p. 1281; G. Alexander and I. Cohen, Measure of $ \pi$'s
 and $ \Lambda$'s emitter radius via Bose-Einstein and \par 
Fermi-Dirac statistics, TAUP 2335-98 (1998).\par
\ndt [2] OPAL Coll., P. D. Acton {\it et al.}, 
Phys. Lett. {\bf B267} (1991) 143;
G. Alexander {\it et al.}, Z. Phys. {\bf C72} (1996) 389.\par
\ndt [3] L3 Coll., L3 98-2270, L3 98-2278, Proc. ICHEP98, Vancouver (1998). \par
\ndt [4] OPAL Coll., P. Acton {\it et al.}, Phys. Lett. {\bf B298} (1993) 456;
R. Akers {\it et al.}, Z. Phys. {\bf C67}(1995) 389.\par
\ndt [5] DELPHI Coll., P. Abreu {\it et al.}, Phys. Lett. {\bf B379} (1996) 
330.\par
\ndt [6] OPAL. Coll., K. Ackerstaff {\it et al.}, 
Euro. Phys. Jou. C5 (1998) 239.\par
\ndt [7] OPAL Coll., G. Abbiendi {\it et al.}, Bose-Einstein Correlations in 
$e^{+} e^{-} \rightarrow W^{+} W^{-}$ at 172 and 183 GeV, \par CERN-EP/98-174 
(1998).\par
\ndt [8] L3 Coll., A two-dimensional study of Bose-Einstein Correlation in Z 
decays al LEP, XXIX Int. Conf. on High En. \par Phys., Vancouver (1998).\par
\ndt [9] OPAL Coll., G. Abbiendi {\it et al.}, 
Intermittency and correlations in hadronic 
decays of the $Z^{0}$, {\bf PN345} (1998); \par E.K.G. Sarkisyan, Study 
of Intermittency and correlations in hadronic $Z^{0}$ decays, 
{\bf CR368}.\par
\ndt [10] OPAL Coll., G. Abbiendi {\it et al.}, A study of spin alignment of
$ \rho (770)^{\pm}$ and $ \omega (782)$ mesons in hadronic $Z^{0}$ decays, \par
OPAL {\bf PN371} (1998).\par
\ndt [11] DELPHI Coll., P. Abreu {\it et al.},
 Phys. Lett. {\bf B406} (1997) 271.\par
\ndt [12] OPAL Coll., K. Ackerstaff {\it et al.}, 
Phys. Lett. {\bf B412} (1997) 210.\par
\ndt [13] MARK III Coll., R. M. Baltrusaitis {\it et al.}, 
Phys. Rev. Lett. {\bf 56}
 (1986) 107.\par
\ndt [14] BES Coll., J. Z. Bai {\it et al.}, 
Phys. Rev. Lett. {\bf 76} (1996) 3502.\par
\ndt [15] G. Forconi, 7th Int. Conf. on Hadron
 Spectroscopy, Brookhaven Nat. Lab. (1997); Available at: \par
http://hpl3sn02.cern.ch/conference/talks97.html.\par
\ndt [16] DM2 Coll., J. E. Augustin {\it et al.}, 
Phys. Rev. Lett. {\bf 60} (1988) 2238.\par
\ndt [17] M. S. Alam {\it et al.}, hep-ex/9805033 (1998).\par
\ndt [18] OPAL Coll., A search for the tensor glueball candidate $f_{J} (2220)$
in hadronic $Z^{0}$ decays, {\bf PN357} (1998).\par

\begin{figure}
\vspace{0.5cm}
\begin{center}
\mbox{
  \epsfig{file=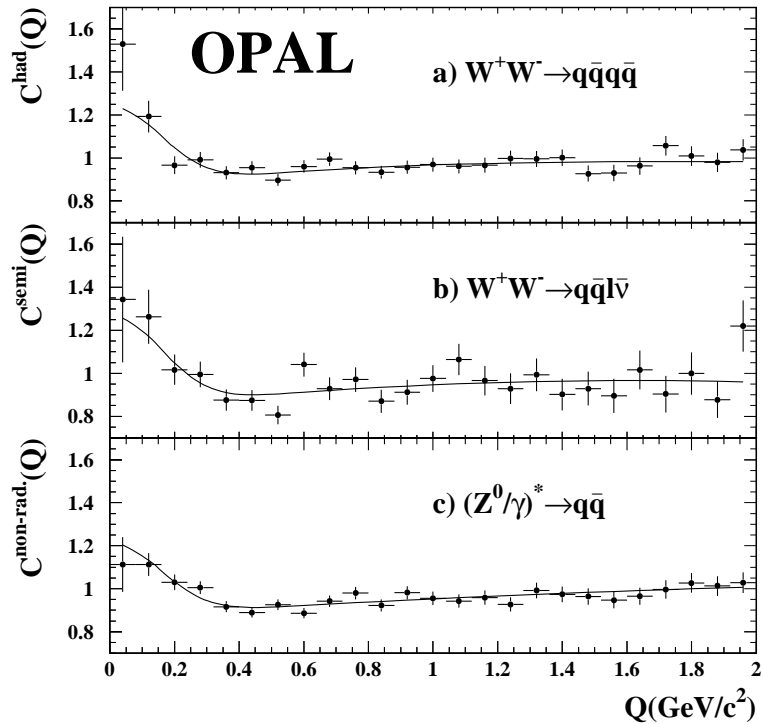,height=9cm}
}
\end{center}
\vspace{0.5cm}
  \caption{ The BEC for like-charge pairs relative to unlike-charge pairs
for three event selections: a) $C^{had} (Q)$ for the fully hadronic, 
b) $C^{semi} (Q)$ for the semileptonic, and c) $C^{non-rad.} (Q)$ for the 
non-radiative events. The Coulomb-corrected data are shown as solid points 
with 
statistical errors. The lines are the result of the simultaneous fit [7].}
  \label{fig:1}
\end{figure}

\begin{figure}
\vspace{0.5cm}
\begin{center}
\mbox{
  \epsfig{file=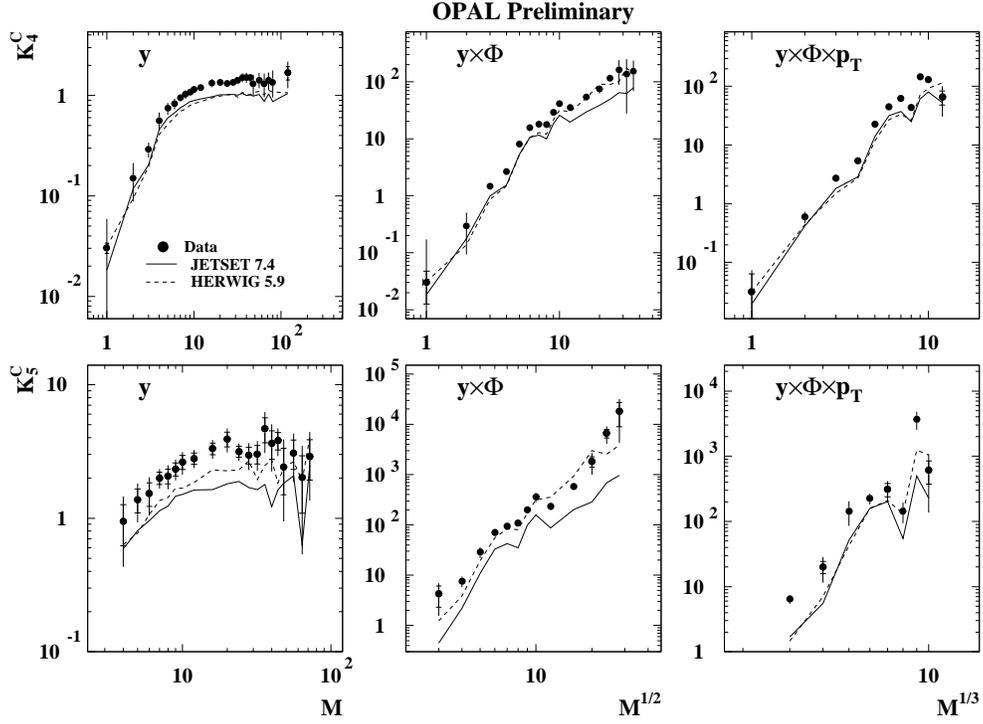,height=9cm}
}
\end{center}
\vspace{0.5cm}
  \caption{ The 4th and 5th FC in one, two, and three dimensions of 
rapidity, azimuthal angle and transverse momentum vs the number of bins 
$M$ [9].}
  \label{fig:2}
\end{figure}

\begin{figure}
\begin{center}
\mbox{
 \epsfig{file=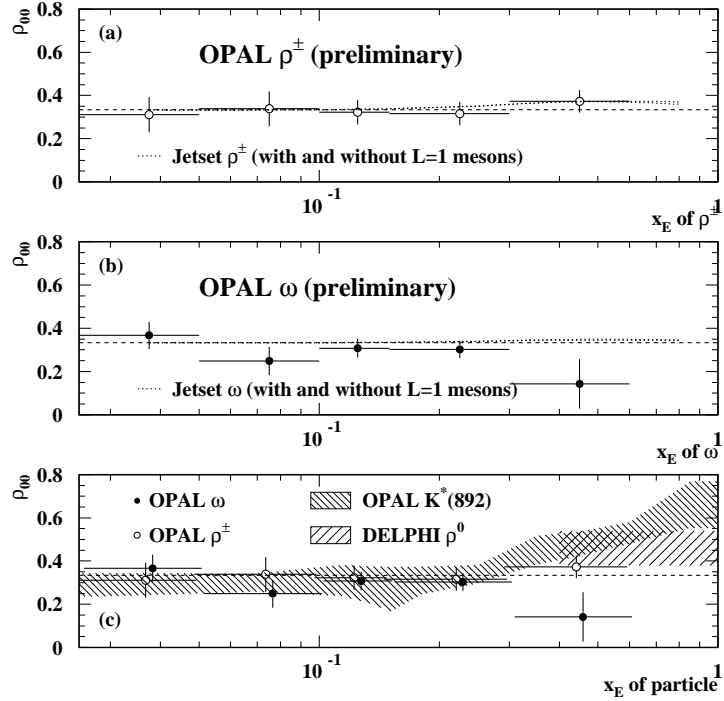,height=9cm}
}
\end{center}
\vspace{0.5cm}
  \caption{\small Measured $ \rho_{00}$ values vs $ x_{E}$ for a) 
$ \rho ^{\pm}$ mesons and b) $ \omega$ mesons produced in $Z^{0}$ decays [10]. 
The dotted lines are the $ \rho_{00}$ values calculated assuming that
$ J^{P}=0^{-} \rightarrow 0^{-}+1^{-}$ decays are the only source of alignment,
and that the number of these decays per hadronic $Z^{0}$ decays is as predicted
by the Monte Carlo. In c) our data are compared with other measurements.
}
  \label{fig:3}
\end{figure}

\begin{figure}
\vspace{0.5cm}
\begin{center}
\mbox{
  \epsfig{file=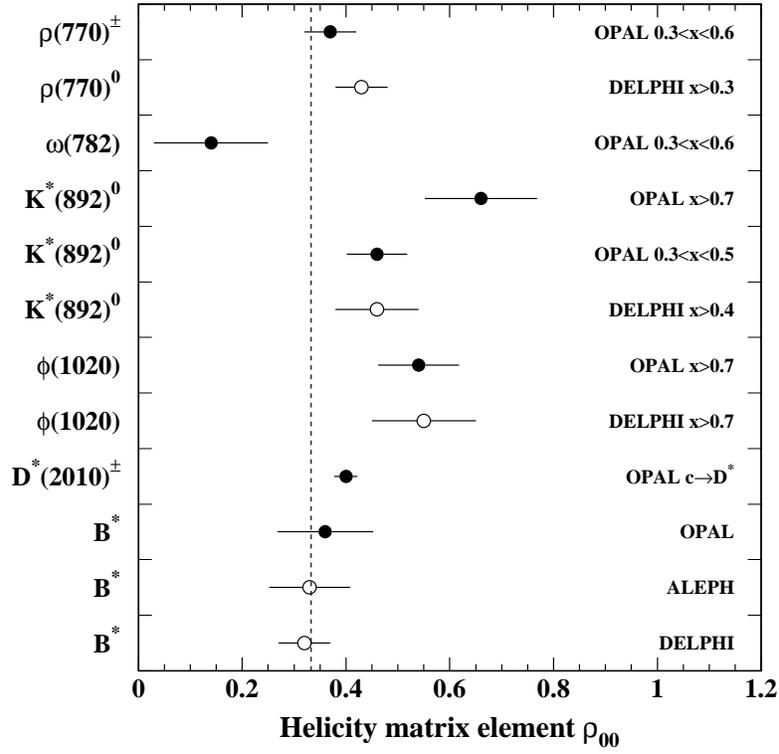,height=10cm}
}
\end{center}
\vspace{0.5cm}
  \caption{ Summary of $ \rho_{00}$ measurements for vector mesons 
produced in $Z^{0}$ decays [10].}
  \label{fig:4}
\end{figure}

\begin{figure}
\begin{center}
\mbox{
  \epsfig{file=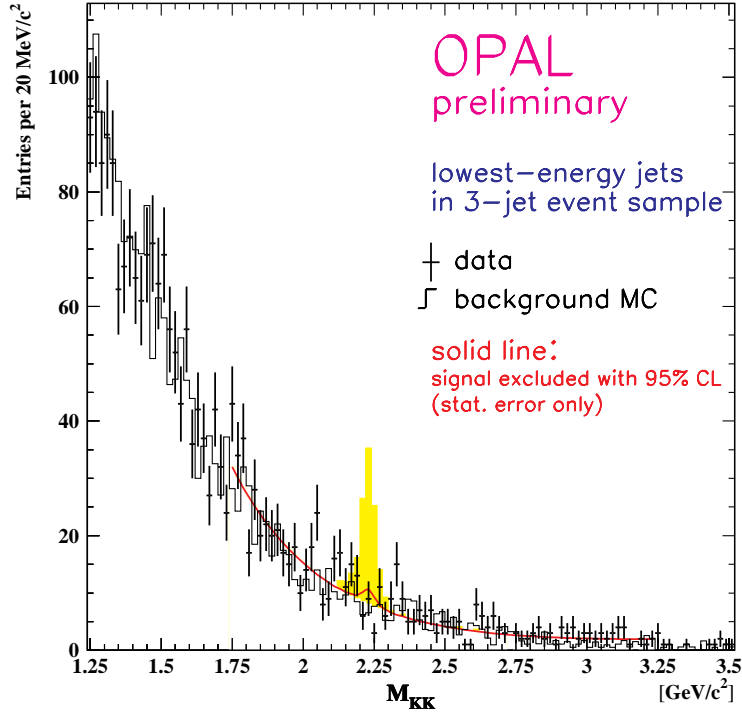,height=9cm}
}
\end{center}
\vspace{0.5cm}
  \caption{Invariant mass spectrum of $ K_{s}^{0}K_{s}^{0}$ pairs for 
the lowest-energy jet in the 3-jet event sample [17]. The Jetset 7.4 
prediction is shown by the open histogram while the lines with error bars 
are the data. The solid line shows the result of the fit with the $95 \% $ 
confidence upper limit for the signal. The shaded histogram is the 
expectation based on the preliminary L3 result.}
  \label{fig:5}
\end{figure}   
 \end{document}